\documentclass[prb,twocolumn,showpacs]{revtex4}
\usepackage{amsmath}
\usepackage{amssymb}
\usepackage{bm}
\usepackage{epsfig}
\usepackage{graphicx}
\usepackage{color}
\usepackage[T2A]{fontenc}
\usepackage{enumitem}

\renewcommand{\Re}{\mathop{\rm Re}}
\renewcommand{\Im}{\mathop{\rm Im}}

\newcount\timehh  \newcount\timemm
\timehh=\time \divide\timehh by 60
\timemm=\time
\begin{document}

\title{Spin noise in quantum dot ensembles}

\author{M.M. Glazov, E.L. Ivchenko}

\affiliation{Ioffe Physical-Technical Institute of the RAS, 194021
  St.-Petersburg, Russia}


\begin{abstract}
We study theoretically spin fluctuations of resident electrons or
holes in singly charged quantum dots. The effects of external magnetic
field and effective fields caused by the interaction
of electron and nuclei spins are analyzed. The fluctuations of spin
Faraday, Kerr and ellipticity signals revealing the spin noise of resident
charge carriers are calculated for the continuous wave probing at the
singlet trion resonance. 
\end{abstract}
\maketitle

\section{Introduction}

Spin noise technique has recently become one of the most promising
methods to study electron spin dynamics in various material
systems,\cite{Mueller2010} including atomic gases~\cite{Crooker_Noise}
and
semicoductors.\cite{Oestreich_noise,muller:206601,PhysRevB.79.035208,crooker2010,dahbashi2012,crooker2012} This
technique first implemented in Ref.~\onlinecite{aleksandrov81} to
observe the magnetic resonance of the sodium atoms is based on the
monitoring of spin fluctuations by means of 
spin Faraday or Kerr rotation effect for the linearly polarized
continuous wave (\emph{cw}) probe. Namely, the spin Faraday
($\vartheta_{\mathcal F}$) or Kerr ($\vartheta_{\mathcal K}$) rotation
angles are 
proportional to the vector component $S_z$ of the instant magnetization of the
medium onto the radiation propagation direction $z$. Hence, the
angle fluctuations reveal the spin autocorrelations
\begin{equation} \label{noise}
\langle \vartheta_{\mathcal F}(t) \vartheta_{\mathcal F}(t')\rangle, \ \langle
\vartheta_{\mathcal K}(t) \vartheta_{\mathcal K}(t')\rangle  \propto \langle S_z(t) S_z(t') \rangle\:,
\end{equation}
where the angle brackets mean averaging over the time $t$ for a fixed
value of the difference $t'-t$. 
The average $\langle S_z(t) S_z(t') \rangle$ characterizes the
magnitude of electron spin fluctuations and contains an important 
information about the spin relaxation and decoherence processes. The
spin noise technique is especially well suited to study slow spin
relaxation in semiconductor
nanostructures.\cite{Mueller2010,dahbashi2012,crooker2012,sherman} 

Contemporary studies of the spin noise in the quantum dot
ensembles made it possible to extract electron and
hole Land\'{e} factors and decoherence
rates.\cite{crooker2010,dahbashi2012} Such 
systems are highly perspective for the future spintronics applications
due to a number of fascinating phenomena, e.g., spin precession
mode-locking where a macroscopic number of spins can precess
synchronously under the conditions of pulse-train optical
excitation.\cite{A.Greilich07212006,A.Greilich09282007} Electron
spin dynamics in quantum dot ensembles is being extensively studied, see
the review articles~[\onlinecite{yakovlev_bayer}],
[\onlinecite{glazov:review}] and 
references therein, while the microscopic theory of the spin noise in
these systems is 
absent to the best of our knowledge. The present paper is aimed to fill the
gap.

Here we address theoretically spin fluctuations of electrons and holes
in quantum dot ensembles. The effects of external magnetic field
as well as the role of hyperfine
coupling of carrier spins with nuclei are discussed in detail. The
features of the spin noise spectra related to the system
inhomogeneity are of special attention. Finally, we derive microscopic
expressions 
for the fluctuation spectra of spin Faraday, Kerr and ellipticity effects for
quantum dot ensembles and perform comparative analysis of these spectra.

\section{Model}\label{sec:model}

The spin fluctuation $\delta {\bm s}(t)$ can be described by the
Langevin method applied to the Bloch equation as follows 
\begin{equation} \label{field}
\frac{\partial \delta {\bm s}(t)}{\partial t} + \frac{ \delta {\bm s}(t)}{\tau_s}  + \delta{\bm s}(t) \times \left({\bm \Omega}_{\bm B} + {\bm \Omega}_N \right)  = {\bm \xi}(t)\:.
\end{equation}
Here $\tau_s$ is the electron spin relaxation time caused
by, e.g., electron-phonon interaction,~\cite{PhysRevB.64.125316,PhysRevB.66.161318} 
${\bm \Omega}_{\bm B}$ and ${\bm \Omega}_N$ are the Larmor precession frequencies 
related to the external magnetic field ${\bm B}$ and the effective
field caused by the hyperfine electron-nuclear interaction, ${\bm
  \xi}(t)$ is the fictitious random force. For simplicity, we assume
an isotropic symmetry of the spin system characterized by single
spin-relaxation time and electron $g$-factor. 
The hyperfine interaction of electron and nuclei spins in the quantum
dot results in the effective magnetic field acting on electron
spin. This field is induced by the nuclear spin fluctuations and
differs from dot to dot, giving rise to the electron spin
dephasing.\cite{kkm_nucl_book,merkulov02,PhysRevLett.88.186802} On
the timescale of electron spin precession in the hyperfine field
induced by the nuclear spin fluctuation,  the latter can
be considered as static.\cite{merkulov02,yugova11}
The magnetic field is assumed to be weak enough in order not to affect the
spin relaxation time $\tau_s$ and nuclear spin fluctuations. Then the
correlator of the Langevin force coincides with that for the
equilibrium spin decoupled from the magnetic field ${\bm B}$ and the
nuclei, namely, 
\begin{equation} \label{correlator_force}
\langle \xi_{\alpha}(t') \xi_{\beta}(t) \rangle = \frac{1}{2 \tau_s} \delta_{\alpha \beta} \delta(t' - t)\:,
\end{equation}
where $\alpha,\beta = x,y$ and $z$ are the
Cartesian coordinates. This equation can readily be derived by using the Langevin approach in the general fluctuation theory~\cite{ggk,LandauStat} applied to a physical variable $x$, e.g., the velocity or spin of a particle, describing by $\dot{x}(t) + x(t)/\tau_0 = 0$ the time decay of its nonequilibrium average value. In this approach the equilibrium fluctuation $\delta x(t)$ satisfies the one-dimensional
equation of random motion $\delta \dot{x}(t) + \delta x(t)/\tau_0 = \xi(t)$ with the inhomogeneous term $\xi(t)$ called the Langevin force. The correlator of $\xi(t)$ is connected with the dispersion 
$\langle \delta x^2 \rangle$ by \cite{LandauStat}
\[
\langle \xi(t') \xi(t) \rangle = (2/\tau_0) \langle \delta x^2 \rangle \delta(t'-t)\:.
\]
Equation (\ref{correlator_force}) follows from this general equation if we take into account that, for the spin $s=1/2$,  the spin-component dispersion $\langle s_{\alpha}^2 \rangle = 1/4$. We stress that the fictitious random force $\bm \xi(t)$ is not related with any real physical processes, it is introduced in Eq.~\eqref{field} in the Langevin approach to provide the proper values of the spin fluctuations in equilibrium. This approach is known as a convenient and effective description of fluctuations.\cite{ggk,lax2}

The spectral decomposition of fluctuations is based on the standard Fourier transforms of the fluctuating spin,
\[
\delta {\bm s}_{\omega} = \int\limits_{- \infty}^{+ \infty} \delta{\bm s}(t) {\rm e}^{{\rm i} \omega t} d t\:,\:
\delta {\bm s}(t)  = \int\limits_{- \infty}^{+ \infty} \delta {\bm s}_{\omega} {\rm e}^{- {\rm i} \omega t} \frac{ d \omega}{2 \pi}\:,
\]
and the spin and random-force correlators
\begin{eqnarray} \label{correlator_spin}
&&(\delta s_{\alpha} \delta s_{\beta})_{\omega} = \int\limits_{- \infty}^{+ \infty} \langle \delta s_{\alpha}(t+ \tau) \delta s_{\beta} (t) \rangle {\rm e}^{{\rm i} \omega \tau} d \tau\:,\\ &&(\xi_{\alpha} \xi_{\beta})_{\omega} = \int\limits_{- \infty}^{+ \infty} \langle \xi_{\alpha}(t+ \tau) \xi_{\beta} (t) \rangle {\rm e}^{{\rm i} \omega \tau} d \tau = \frac{1}{2 \tau_s} \delta_{\alpha \beta}\:.
\nonumber
\end{eqnarray}
The Fourier component $\delta {\bm s}_{\omega}$ satisfies the vector equation obtained from Eq.~(\ref{field})
by the replacements $\partial/ \partial t \to - {\rm i} \omega$,
$\delta {\bm s}(t) \to \delta {\bm s}_{\omega}$ and ${\bm \xi}(t) \to
{\bm \xi}_{\omega}$. According to Eq.~(\ref{correlator_force}) the
double correlator of the random-force Fourier transforms is given by 
\[
\langle \xi_{\alpha, \omega} \xi_{\beta, \omega'}^* \rangle = \delta(\omega - \omega')  \delta_{\alpha \beta} 
\frac{\pi}{\tau_s}\:.
\]

The solution of Eq.~(\ref{field}) for the spin pseudovector reads
\begin{equation} \label{solution}
\delta {\bm s}_{\omega} =  \tau_{s,\omega} \frac{{\bm \xi}_{\omega} + \tau_{s,\omega} {\bm \Omega} \times {\bm \xi}_{\omega} + \tau_{s,\omega}^2 {\bm \Omega} ({\bm \xi}_{\omega} \cdot {\bm \Omega} ) }{1 + {\bm \Omega}^2 \tau^2_{s,\omega}}\:,
\end{equation}
where 
\begin{equation}
\label{Omega:inB}
{\bm \Omega} = {\bm \Omega}_{\bm B} + {\bm \Omega}_N\hspace{2 mm} \mbox{and} \hspace{2 mm}\tau_{s,\omega} = \frac{\tau_s}{1-\mathrm i \omega \tau_s}\:.
\end{equation}
The relation (\ref{solution}) between the spin ${\bm s}_{\omega}$ and the random force ${\bm \xi}_{\omega}$ can formally be 
obtained from the equation (9) in Ref.~\onlinecite{merkulov} relating the average spin $\bar{\bm s}$ and the initial spin ${\bm s}_0$ by replacing $\bar{\bm s}$ to ${\bm s}_{\omega}$, $\tau_c$ to $\tau_{s,\omega}$, and ${\bm s}_0$ to $\tau_{s,\omega} {\bm \xi}_{\omega}$. By introducing the linear-response tensor ${\bm \chi}(\omega)$ defined as
\begin{equation} \label{suscept}
\delta s_{\alpha, \omega} = \chi_{\alpha \beta}(\omega) \xi_{\beta, \omega}\:,
\end{equation}
we can present the spin-fluctuation spectrum in the form 
\begin{eqnarray} \label{fd}
(\delta s_{\alpha} \delta s_{\beta})_{\omega} &=& \frac{1}{2 \tau_s} \sum\limits_{\gamma} \chi_{\alpha \gamma}(\omega) \chi^*_{\beta \gamma}(\omega) \nonumber \\ &=& \frac{1}{2 \tau_s} \left[ {\bm \chi}(\omega) {\bm \chi}^{\dag}(\omega) \right]_{\alpha \beta}\:.
\end{eqnarray}
The components of the tensor ${\bm \chi}(\omega)$ are explicitly given by
\begin{equation} \label{componchi}
\chi_{\alpha \beta}(\omega) = \frac{\tau_{s,\omega} \left( \delta_{\alpha \beta}
- \delta_{\alpha \beta \gamma} \Omega_{\gamma} \tau_{s,\omega} + \tau_{s,\omega}^2 \Omega_{\alpha} \Omega_{\beta} \right)}{1 + {\bm \Omega}^2 \tau^2_{s,\omega}} \:,
\end{equation}
where $\delta_{\alpha \beta \gamma}$ is the unit antisymmetric
third-rank tensor. Equation (\ref{fd}) can also be derived
by means of the fluctuation-dissipation theorem, see Appendix for
details, or from kinetic equations for the spin-spin
  correlation functions $\langle \delta s_{\alpha}(t) \delta
  s_{\beta}(t')\rangle$ similar to the general approach of Ref.~\onlinecite{LandauKin}. To summarize, Eqs.~\eqref{fd} and \eqref{componchi} are valid provided: (i) $\tau_s$ is magnetic field independent, (ii) nuclear field $\bm \Omega_{N}$ is static (or quasistatic).

Let us consider the simple limiting cases to show that Eq.~(\ref{fd})
readily describes them. In the absence of external and internal
fields, $\bm \Omega = 0$, the response reduces to $\chi_{\alpha \beta}(\omega) = \tau_{s,\omega} \delta_{\alpha \beta}$ and we have
\begin{equation} \label{zerofield}
(\delta s_{\alpha} \delta s_{\beta})_{\omega} = \frac{|\tau_{s,\omega}|^2}{2 \tau_s} \delta_{\alpha \beta} =
\frac12 \frac{\tau_s}{1 + (\omega \tau_s)^2} \delta_{\alpha \beta}\:,
\end{equation} 
in agreement with Refs.~\onlinecite{Mueller2010}, \onlinecite{ivchenko73fluct_eng}. For a system subjected to an external magnetic field ${\bm B}$ but 
free from the hyperfine interaction, it is convenient to use the Cartesian coordinate frame $1,2,3$ with the axis $3 \parallel {\bm \Omega}_{\bm B}$. In this case
\begin{eqnarray} \label{chimag}
|| \chi_{\alpha \beta}(\omega)|| = \frac{\tau_{s,\omega}}{1 + (\Omega_3 \tau_{s,\omega})^2 } \hspace{1 cm} \\  \times \left[  \begin{array}{ccc} 1 & - \Omega_3 \tau_{s,\omega} & 0 \\ \Omega_3 \tau_{s,\omega} & 1 & 0 \\ 0 & 0 & 1 + (\Omega_3 \tau_{s,\omega})^2 \end{array} \right] \:. \nonumber
\end{eqnarray} 
Substitution of this tensor into Eq.~(\ref{fd}) leads to the following nonzero components of the tensor
$(\delta s_{\alpha} \delta s_{\beta})_{\omega}$:
\begin{eqnarray} \label{noise:field}
&&(\delta s_3^2 )_{\omega} = \frac{\pi}{2} \Delta(\omega)\:, \\
&&(\delta s_1^2 )_{\omega} = (\delta s_2^2 )_{\omega} = \frac{\pi}{4} [ \Delta(\omega - \Omega_3) + 
\Delta(\omega + \Omega_3)]\:, \nonumber \\
&&(\delta s_2 \delta s_1)_{\omega} = (\delta s_1 \delta s_2)^*_{\omega} = \frac{2 {\rm i} \omega \Omega_3 \tau_s^2}{1 + \tau^2_s (\omega^2
+ \Omega_3^2)}(\delta s_1^2 )_{\omega}\:, \nonumber
\end{eqnarray}
where
\[
\Delta(x) = \frac{1}{\pi} \frac{\tau_s}{ 1 + (x \tau_s)^2}\:.
\]
Equations (\ref{noise:field}) are also valid in the absence of an
external magnetic field but in the presence of a fixed hyperfine
field, in this case $3 \parallel \bm \Omega_{N}$. 

\section{Spin fluctuations in quantum-dot
  ensembles}\label{sec:ensemble} 

Until now, we described electron spin fluctuations in a single dot. In a quantum-dot ensemble the spin noise spectrum per quantum dot is obtained by averaging Eq.~(\ref{fd}) over the direction and absolute value of ${\bm \Omega}_N$
\begin{equation}
\label{general}
(\delta s_{\alpha} \delta s_{\beta})_{\omega} = \frac{1}{2 \tau_s} \int d {\bm \Omega}_N {\cal F}({\bm \Omega}_N) \left[ {\bm \chi}(\omega) {\bm \chi}^{\dag}(\omega) \right]_{\alpha \beta}\:,
\end{equation}
where ${\cal F}({\bm \Omega}_N)$ is the distribution function of the nuclear fields acting on electron spins in the quantum dot ensemble. Due to the $s$-type character of the conduction band Bloch functions the
hyperfine interaction strength is proportional to the scalar product
of electron and nuclear spins, resulting in an isotropic distribution
of the nuclei-induced electron spin precession frequencies ${\bm
\Omega}_N$. Therefore $\mathcal F({\bm \Omega}_N)$ depends only on the
absolute value of the nuclear field. 
\subsection{Spin fluctuations in the absence of external field}\label{subsec:hf}
At zero magnetic field, the averaging over the direction results in the simplification of $(\delta s_{\alpha} \delta s_{\beta})_{\omega}$ to $\delta_{\alpha \beta} (\delta s_{\alpha}^2)_{\omega}$ with $(\delta s_{\alpha}^2)_{\omega} = [(\delta s_1^2)_{\omega} + (\delta s_2^2 )_{\omega} + (\delta s_3^2 )_{\omega}]/3$ and, therefore,
\begin{eqnarray}
\label{electron:nuclei}
(\delta s_{\alpha}^2)_{\omega} =  \frac{\pi}{6} \left\{
  \Delta(\omega) \right. \hspace{ 2 cm}\\  +  \int_{0}^{\infty} d \Omega_N \ F(\Omega_N) \
[\Delta(\omega - \Omega_N) + \Delta(\omega + \Omega_N)] \}\:, \nonumber
\end{eqnarray} 
where $$F(\Omega_N) = 4\pi \Omega_N^2\mathcal F(\Omega_N)$$ is the
distribution of absolute values of the hyperfine fields. The
factor $4\pi \Omega_N^2$ takes into account a three-dimensional
character of the random vector $\bm \Omega_N$.
Equation~(\ref{electron:nuclei}) clearly shows that the spin noise
spectrum contains two contributions. First one is centered at
$\omega=0$ and stems from fluctuations of the spin component directed
along the nuclear field which is considered as static
 here. In that way the zero-frequency peak bears information
about the single-electron spin relaxation time $\tau_s$.  Note, that
for sufficiently long spin relaxation times the electron hyperfine
interaction with nuclei may modify slow electron spin dynamics and,
  correspondingly,  the low frequency spin noise spectra, due to coupled spin dynamics of
  electrons and nuclei.\cite{coupled1}
The second
contribution to the spin noise spectrum reflects electron spin
precession in random nuclei-induced fields. Provided that
$\Omega_N \tau_s \gg 1$ this contribution to
the spin noise spectrum, for $\omega>0$, reduces to $(\pi/6)
F(\omega)$  and describes the distribution of the nuclear-induced 
spin precession fluctuations.

\begin{figure}[hptb]
\includegraphics[width=0.45\textwidth]{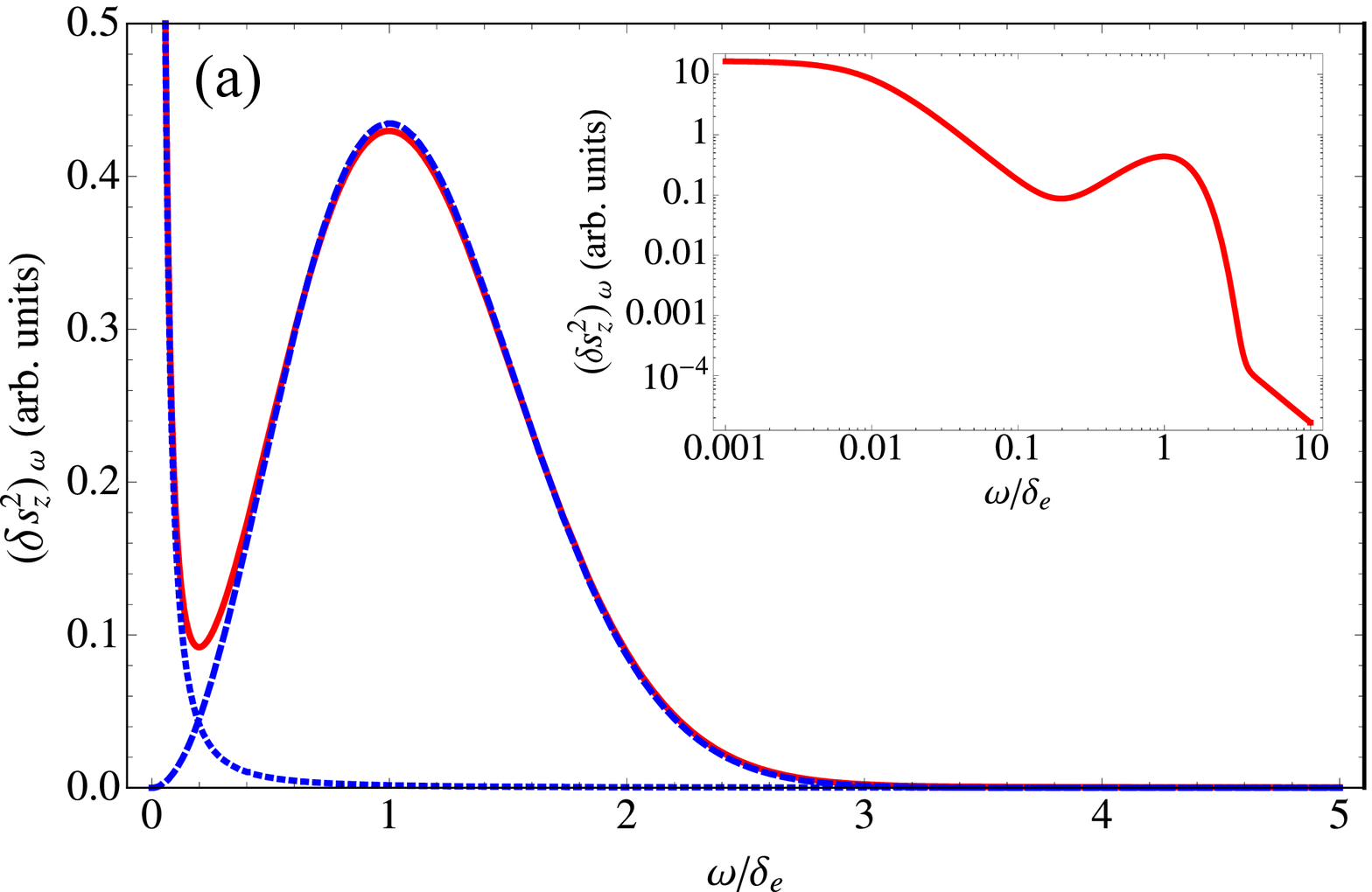}\\
\includegraphics[width=0.45\textwidth]{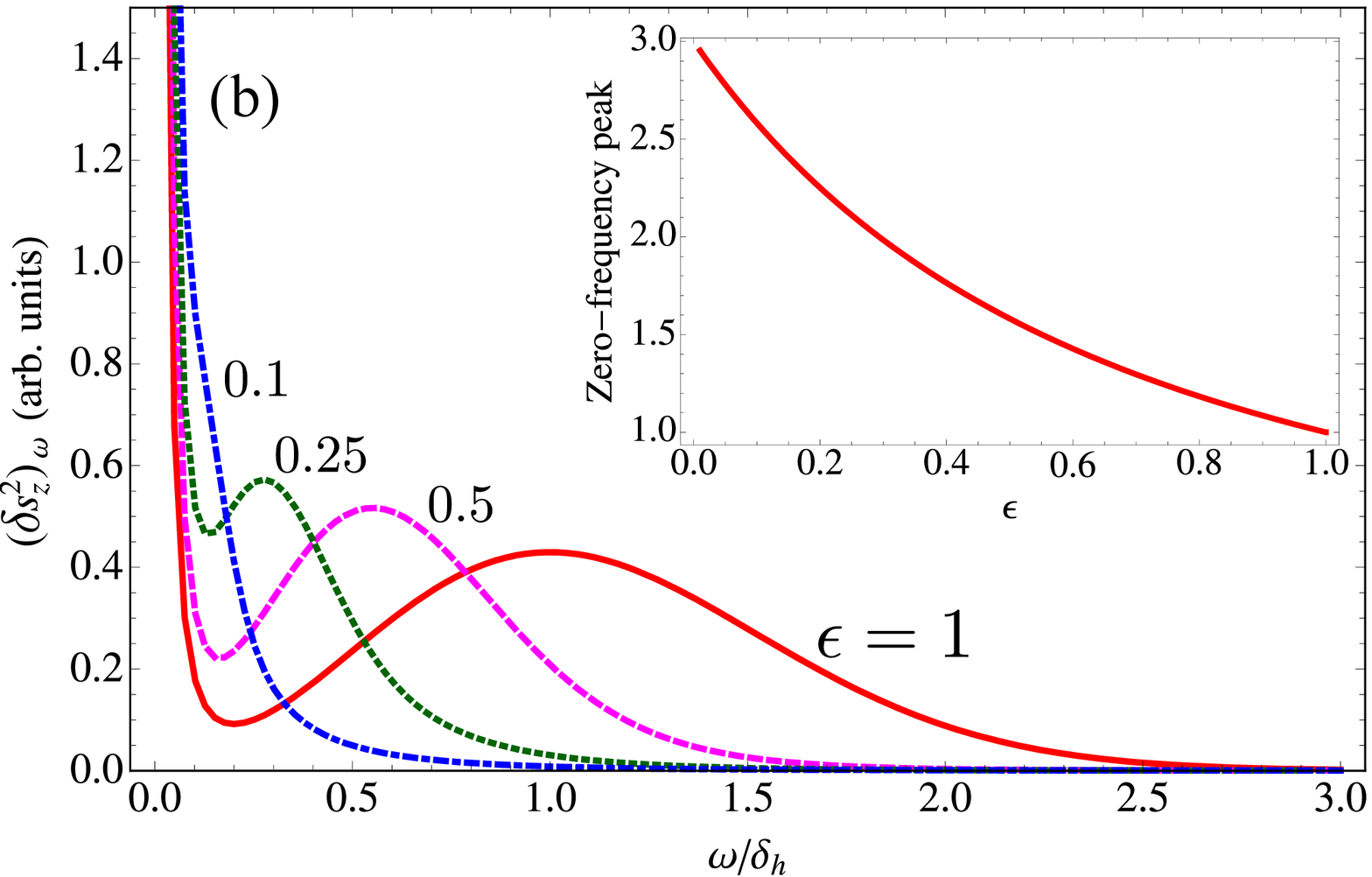}
\caption{(a) Spin noise spectrum for electrons (red solid curve) and
its decomposition into the zero frequency peak (dotted) and the peak corresponding
to nuclei-induced precession (dashed) calculated for  $\delta_e
\tau_s=100$. Inset shows the spin noise 
spectrum in the double logarithmic scale. (b)
Heavy-hole pseudospin noise spectrum calculated for different values of anisotropy parameter
$\epsilon$ (marked at each curve). Inset shows the amplitude of
zero-frequency peak (normalized to its value at $\epsilon=1$) as a
function of the anisotropy parameter calculated for $\delta_h \tau_s=100$.}
\label{fig:sns:zero}
\end{figure}

The electron spin noise spectrum calculated for the quantum dot
ensemble is shown in Fig.~\ref{fig:sns:zero}(a) for the Gaussian
distribution of nuclear fields acting on the electron spin, $\mathcal
F(\bm \Omega_N) =
(\sqrt{\pi}\delta_e)^{-3} \exp{(-\Omega_N^2/\delta_e^2)}$, where $\delta_e$
describes the dispersion of the nuclear field fluctuation.\cite{merkulov02,chemphys} Two
contributions to the spin noise spectrum are clearly seen: the narrow
peak at $\omega=0$, which is well described by $\Delta(\omega)$, and
much wider peak at $\omega =\delta_e$ corresponding to the distribution
of nuclear fields $(\pi/6) F(\omega)$. The Fourier transform of
Eq.~(\ref{electron:nuclei}) to the time domain gives the relaxation
dynamics of the spin component $s_\alpha$; for the Gaussian distribution of nuclear spins
and in the limit $\tau_s \to \infty$, it
  reduces to Eq. (10) of Ref.~\onlinecite{merkulov02}, see also
  Ref.~\onlinecite{zhang06}. 

Next we turn to the spin fluctuations of holes in quantum dots.  Here, for distinctness, we
consider heavy-hole states, namely, the doubly-degenerate hole states with
the angular momentum projection $J_z= \pm 3/2$ onto the growth axis of
the quantum dot. Occupation of this pair of states can be described by means of
the pseudospin-$1/2$ three-component vector $\tilde{\bm s}$. 
Contrary to electrons, the coupling of a heavy hole and nuclear
spins results not from the contact hyperfine interaction but rather from the relatively weak dipole-dipole interaction.
The latter is strongly anisotropic in quantum dots.
In particular, if the mixing of heavy-hole ($\pm 3/2$) and light-hole ($\pm 1/2$) states and contributions related with the cubic
symmetry of the crystalline lattice are disregarded, the effective
magnetic field experienced by the hole spin is parallel to the $z$ axis
and proportional to the nuclear spin $z$
component.\cite{perel76eng,PhysRevB.78.155329,PhysRevB.79.195440}  In
real quantum-dot structures heavy holes additionally experience an effective field
of the in-plane nuclear-spin components~\cite{PhysRevLett.102.146601} which
is, however, weaker than that caused by the $z$ component. As
a result, we model the distribution of effective fields acting on the
hole spins by an anisotropic Gaussian function:
\begin{equation}
\mathcal F_h(\bm \Omega) =
\frac{1}{\pi^{3/2} \delta_h^3 \epsilon^2} \exp{\left(-\frac{\Omega_z^2}{\delta_h^2} - \frac{\Omega_\perp^2}{(\epsilon\delta_h)^2}\right)}\:,
\end{equation}
where $\bm \Omega_{\perp} = (\Omega_x,\Omega_y)$ is the in-plane
component of the effective Larmor frequency,  a value of $\epsilon$
lies in the interval between 0 and 1 and characterizes the relative
strength of the hole coupling with the in-plane nuclear fields, and an
in-plane anisotropy of hole-nuclear coupling is neglected.
Fluctuations of the $z$-component of hole pseudospin are given by 
\begin{multline}
\label{hole:nuclei}
( \delta \tilde s_z^2)_{\omega} =  \frac{\pi}{2} 
  \int d \bm \Omega \  \mathcal F_h(\bm \Omega) \\ 
\times
\left\{ \frac{\Omega_z^2}{\Omega^2} \ \Delta(\omega) 
+
  \frac{\Omega_\perp^2}{\Omega^2} \
\biggl[\Delta(\omega - \Omega) + \Delta(\omega + \Omega)\biggr] \right\}\:,
\end{multline} 
where $\bm \Omega = (\Omega_x,\Omega_y,\Omega_z)$ and $\Omega^2 = \Omega_\perp^2 + \Omega_z^2$.

Figure~\ref{fig:sns:zero}(b) shows the heavy-hole pseudospin noise spectrum 
calculated for different values of the anisotropy
parameter $\epsilon$. The fluctuation spectrum is similar to
that for electrons, see curve corresponding to $\epsilon=1$ and
Fig.~\ref{fig:sns:zero}(a), and contains two peaks: at zero frequency,
due to the first term in Eq.~(\ref{hole:nuclei}), and the
high-frequency peak described by the second term in the curly brackets of
Eq.~(\ref{hole:nuclei}). Note that for $\epsilon < 1$ the
height of the zero-frequency peak is enhanced as compared with that
for electrons, see inset in Fig.~\ref{fig:sns:zero}(b). In the
limit of strong anisotropy $\epsilon \ll 1$ it is three times higher
than in the isotropic case, since, for $\epsilon \to 0$, the effective nuclear field is always
directed along $z$ axis. The high-frequency peak shifts towards the zero frequency with decreasing
$\epsilon$ and eventually merges with the zero-frequency peak. Note,
that while comparing quantitatively with experiments one has to allow
for the hole spin relaxation anisotropy and introduce the longitudinal
and transverse relaxation times.

\subsection{Spin fluctuations in the external magnetic
  field}\label{sec:b}

In the presence of an external magnetic field the electron (hole)
spin precession frequency $\bm \Omega$ has, according to
Eq.~(\ref{Omega:inB}), two contributions due to (i) interaction with
nuclei $\bm \Omega_N$ and (ii) interaction with an  external field,
$\bm \Omega_{\bm B}$. Making
use of Eq.~(\ref{general}) we recast the spectrum of the spin $z$ component
fluctuations in the following form [cf. Eq.~(\ref{hole:nuclei})]
\begin{align}
\label{exact}
&( \delta s^2_z)_{\omega} = \frac{\tau_s}{2} \int d{\bm
\Omega}_N \ \mathcal F({\bm \Omega}_N) \times \\
& \left\{\sin^2{\theta} \frac{1+(\omega^2+\Omega^2)\tau_s^2}{[1 + (\omega - \Omega)^2\tau_s^2][1 + (\omega + \Omega)^2\tau_s^2]}  + \frac{ \cos^2{\theta}}{1 + \omega^2\tau_s^2}  \right\}.\nonumber
\end{align}
where $\theta$ is the angle between the vector ${\bm \Omega} =
{\bm \Omega}_N + {\bm \Omega}_{\bm B}$ and $z$ axis
and $\Omega = |{\bm \Omega}_{\bm B} + {\bm \Omega}_N|$. In the absence
of the nuclear spin fluctuations Eq.~(\ref{exact}) reduces (up to a
common factor) to Eq. (5) of Ref.~\onlinecite{braun}.

\begin{figure}[t]
\includegraphics[width=0.45\textwidth]{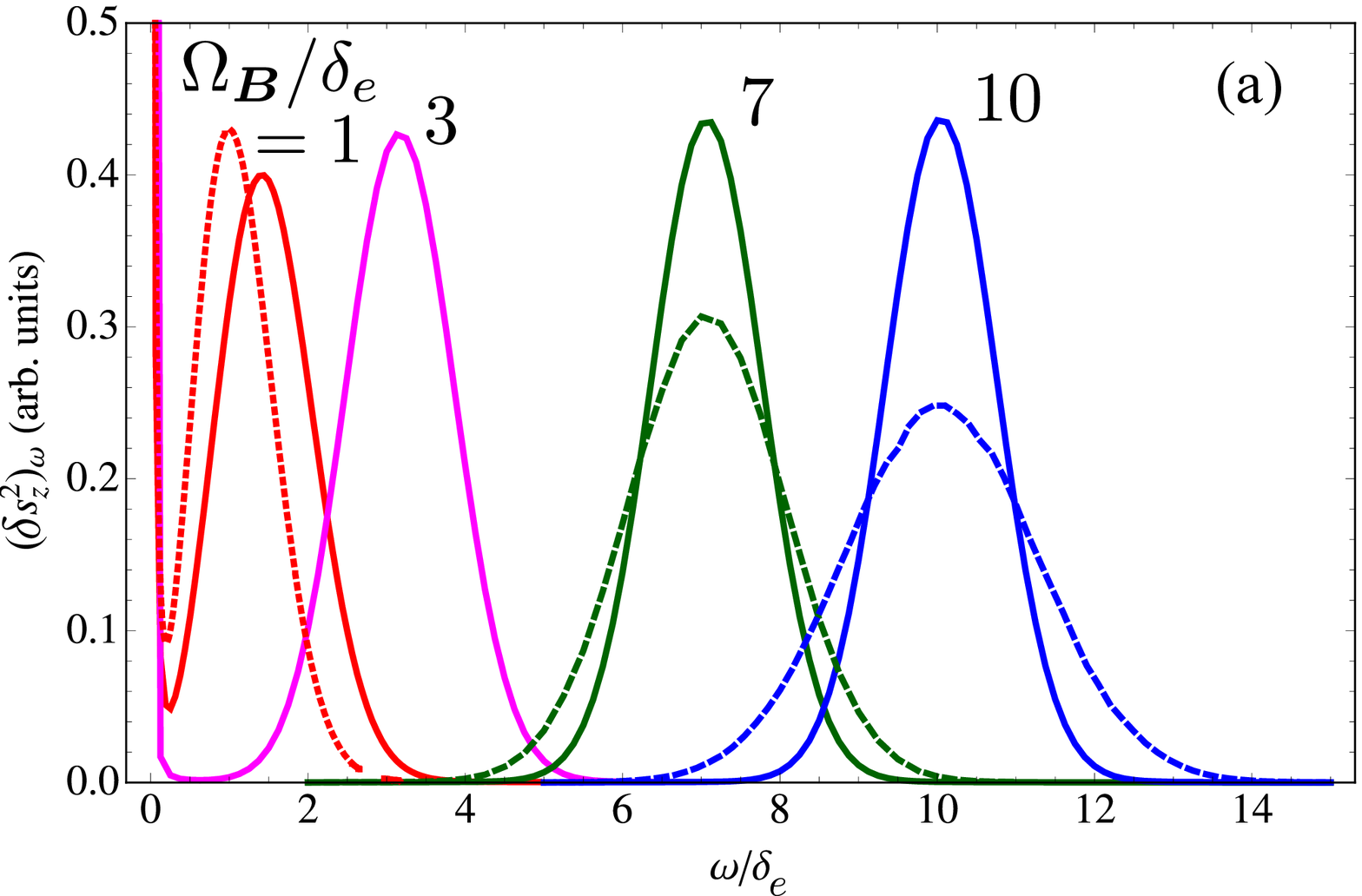}\\
\includegraphics[width=0.45\textwidth]{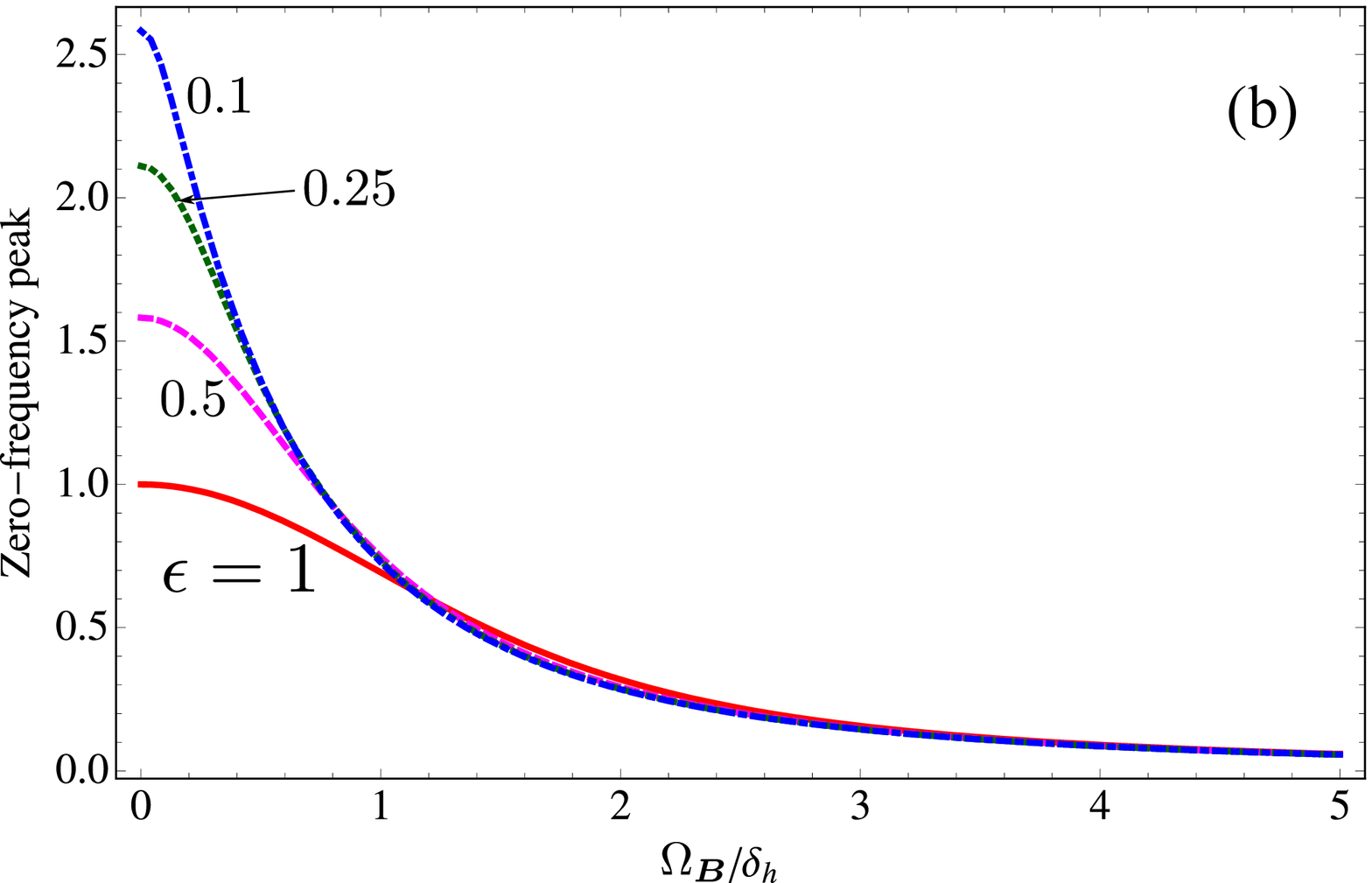}
\caption{(a) Spin noise spectrum for electrons in the transverse magnetic
  field (solid) for $\delta_e \tau_s=100$. Ratio between $\Omega_{\bm B}$ and $\delta_e$ is indicated at each
  curve. Red dotted curve shows electron spin
  noise spectrum at $\bm B=0$ [c.f. red curve in
  Fig.~\ref{fig:sns:zero}(a)], dashed green and blue curves demonstrate
spin noise spectra calculated with a $10$\% spread of
electron $g$-factor. (b) Zero-frequency peak amplitude (normalized to
its value at $\bm B=0$, $\epsilon=1$) calculated as a function of the
magnetic field for $p$-type quantum dot ensemble. Different curves
correspond to different values of 
the anisotropy parameter $\epsilon$ (marked at each curve).}
\label{fig:sns:nonzero}
\end{figure}

Figure~\ref{fig:sns:nonzero}(a) shows the electron spin noise spectrum
in the transverse magnetic field $\bm \Omega_{\bm B} \perp z$. Such a
case corresponds to the most widespread experimental configuration of
the spin noise measurements.~\cite{Mueller2010} Calculation shows
that the noise peak at $\omega \ne 0$ shifts proportionally to $\Omega_{\bm
  B}$ for sufficiently high magnetic fields ($\Omega_{\bm B} \gg
\delta_e$) and its height slightly increases with an increase of
magnetic field. In such a case, the spin
fluctuation dispersion is controlled only by the parallel to
$\bm\Omega_{\bm B}$ component of the nuclear spin fluctuations and,
at $\delta_e\tau_s \gg 1$, the spin noise spectrum can be reduced to 
\begin{equation}
\label{parallel:fluct}
(\delta s_z^2)_\omega = \frac{\pi}{8} \int\limits_0^\infty
  \frac{
F\left[\sqrt{\Omega_N^2 + (\omega-\Omega_{\bm B})^2}\right]    \Omega_N d\Omega_N} 
{\Omega_N^2+(\omega-\Omega_{\bm B})^2} \,
  . 
\end{equation}
With the further increase in magnetic field, the spin noise spectrum
is additionally 
affected by the inhomogeneous broadening of electron $g$-factor values
similarly to the 
spin dephasing of the localized carriers.\cite{yakovlev_bayer} This
effect is illustrated by the dashed curves in
Fig.~\ref{fig:sns:nonzero}(a) where a $10$\% spread of
electron $g$-factor values was additionally considered. The $g$-factor dispersion
is modeled by the Gaussian distribution, see Ref.~\onlinecite{glazov08a}
for details. The spin noise spectra of quantum dots with resident
holes have a similar shape.

The transverse magnetic field affects also the zero-frequency
peak. Figure~\ref{fig:sns:nonzero}(b) demonstrates the magnitude of
the zero-frequency peak as a function of magnetic field calculated
for the case of the positively charged quantum dot ensemble. Different
curves correspond to 
different values of anisotropy parameter $\epsilon$. Figure shows that
the magnetic field suppresses the zero-frequency peak, because the higher
the field, the smaller the probability to find the quantum dot with
$\bm \Omega = \bm \Omega_N + \bm \Omega_{\bm B} \parallel z$. In other
words, the average value of 
$\cos^2{\theta}$ in Eq.~(\ref{exact}) decreases with an increase in
the transverse magnetic field.

\begin{figure}[hptb]
\includegraphics[width=0.49\textwidth]{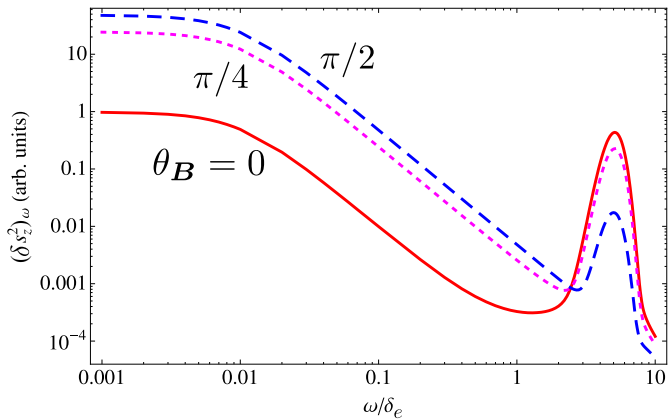}
\caption{Spin noise spectrum for electrons in the magnetic field $\bm
  B$ applied at the angle $\theta_{\bm B}=0$ (red solid), $\pi/4$ (magenta
  dotted) and $\pi/2$ (blue dashed) to the $(xy)$
  plane. Electron $g$-factor is assumed to be isotropic, $\Omega_{\bm
    B}/\delta_e=5$, $\delta_e\tau_s=100$.}
\label{fig:sns:long}
\end{figure}

The situation is different if the applied magnetic field acquires a
longitudinal (parallel to $z$-axis) component. The spin noise spectra
are presented in Fig.~\ref{fig:sns:long}. It is clearly seen that the
peak corresponding to the non-zero frequency becomes suppressed with
the increasing of the tilt angle of magnetic field with respect to the
quantum dot plane, while the zero-frequency peak becomes
higher. Indeed, the considerable $z$ component of the external
magnetic field leads to the diminishing of the transversal
nuclear field role resulting in an enhancement of the
zero-frequency peak, in agreement with experiment.\cite{dahbashi2012}
It is worth to mention that the total spin fluctuation $\sum_{\alpha} (\delta s^2_{\alpha})_{\omega}$ 
is independent of $\theta_{\bm B}$.

\section{Manifestations of the spin noise in Faraday, Kerr and
  ellipticity effects}\label{sec:signals}

Here we analyze the fluctuations of the spin Faraday, Kerr and
ellipticity effects detected by the \emph{cw} linearly
polarized probe beam propagating along the growth axis of the quantum
dot ensemble. We assume that the probe frequency $\omega_{\rm pr}$ is
close to the singlet trion resonance in the quantum dot and
consider, as an example, an ensemble of $n$-type singly charged quantum
dots.

If all dots in the ensemble are identical, i.e., have the same trion
resonance frequency $\omega_0$ and the same electron spin fluctuation property,
the Faraday ($\vartheta_{\mathcal F}$) and ellipticity 
($\vartheta_{\mathcal E}$) angles read
\begin{equation}
\label{spinFE}
\vartheta_{\mathcal E}(t) + \mathrm i \vartheta_{\mathcal F}(t) = \frac{\delta s_{j,z}(t)}{S} \frac{3\pi}{4 q^2}  G_0 (\omega_0 - \omega_{\rm pr}),
\end{equation}
where $S$ is the sample area, $\delta s_{j,z}$ is the spin fluctuation in $j$th dot, $q =
\sqrt{\varepsilon_b} \omega_0/c$ is the wavevector of the 
electromagnetic wave in the sample (the background dielectric constant
of both the quantum dots and the matrix are assumed to be the same and
equal to $\varepsilon_b$), and the function  
\begin{equation}
\label{G}
G_0(\Lambda)=\mathrm i
\Gamma_0/(\Lambda + \mathrm i 
\Gamma_0)
\end{equation} 
describes the spin signal sensitivity at the continuous wave
probing,
$\Gamma_0$ is the trion radiative decay rate, any non-radiative losses are
neglected.  Equation~(\ref{spinFE}) directly follows from the
definition of the spin Faraday and ellipticity signals for the quantum
dot ensembles, see Refs.~\onlinecite{glazov:review}, \onlinecite{yugova09} for details. 
The Kerr rotation angle measured in the reflection geometry is
determined by the phase acquired by the probe pulse in the cap layer
of the structure, it is proportional to a certain linear combination of
$\vartheta_{\mathcal E}$ and~$\vartheta_{\mathcal F}$.\cite{yugova09}

Temporal fluctuations and frequency dispersion of Faraday, ellipticity
and Kerr signals can be 
calculated by means of Eq.~(\ref{spinFE}). It is
important to stress that electron spins in different quantum dots of the
ensemble are uncorrelated. As a result, the spectra of Faraday
rotation and ellipticity fluctuations for an ensemble of identical
dots read:
\[
( \vartheta_{\mathcal F}^2 )_\omega =  (\delta s_z^2)_\omega  \left(\frac{3\pi}{4 q^2}\right)^2 \frac{N_{QD}^{2D}M}{S}\frac{(\omega_0-\omega_{\rm
    pr})^2\Gamma_0^2} {[(\omega_0-\omega_{\rm pr})^2+\Gamma_0^2]^2},
\]
\[
( \vartheta_{\mathcal E}^2 )_\omega =  (\delta s_z^2)_\omega  \left(\frac{3\pi}{4 q^2}\right)^2 \frac{N_{QD}^{2D}M}{S} \frac{\Gamma_0^4} {[(\omega_0-\omega_{\rm pr})^2+\Gamma_0^2]^2}.
\]
Here $N_{QD}^{2D}$ is the two-dimensional quantum dot density in the
layer and $M$ is the number of layers.
These equations should be averaged over all possible
resonance frequencies $\omega_0$ and electron spin precession
frequencies $\bm \Omega$ in the
ensemble. To that end, we consider two important limits: (\emph{a}) electron
spin precession frequencies and optical transition frequencies are not
correlated at all, and (\emph{b}) electron spin precession frequency is a
certain function of the optical transition energy. The case (\emph{a}) can be
realized in relatively small external magnetic fields, where the
nuclear spin fluctuations determine the electron spin precession, and
the case (\emph{b}) may be important in rather high magnetic fields where
nuclear effects are negligible, in this case the link between the spin
precession frequency and optical transition frequency results from the
dependence of electron $g$-factor on the band gap of the
nanosystem.\cite{ivchenko05a}

If (i) spread of the quantum dot
resonance frequencies is much broader than $\Gamma_0$ (this condition
holds for the self-organized quantum dot ensembles studied in
Ref.~\onlinecite{crooker2010}), and (ii) the  probe frequency is not too
close to the edges 
of the quantum dot distribution,
then, in case (\emph{a}), the fluctuation spectra of the Faraday and
ellipticity signals are simply proportional to $(s_z^2)_\omega$ and
weakly depend on the probe (optical) 
frequency. Under the above conditions the magnitudes of the Faraday
and ellipticity 
fluctuations coincide.

In the limiting case (\emph{b}), the nuclear fluctuations can be
disregarded and electron spin precession frequency $\Omega \equiv
\Omega_{\bm B}$ is well described by a linear function of the
optical transition frequency $\omega_0$:\cite{yakovlev_bayer}
\begin{equation}
\label{AC}
\Omega_{B}(\omega_0) = A\omega_0 +C,
\end{equation}
where $A$ and $C$ are constants. Under
condition $\Omega_{\bm B} \tau_s \gg 1$, the function $\Delta(x)$ in
Eqs.~(\ref{noise:field}) reduces to the Dirac delta-function and
fluctuation spectra of 
Faraday and ellipticity signals are proportional to 
\[
(\vartheta_{\mathcal F}^2)_\omega \propto \left\{\Im{\left[G_0\left(\frac{\omega-C}{A} - \omega_{\rm
          pr}\right)\right]}\right\}^2,
\]
\[
(\vartheta_{\mathcal E}^2)_\omega \propto  \left\{\Re{\left[G_0\left(\frac{\omega-C}{A}  - \omega_{\rm pr}\right)\right]}\right\}^2,
\]
respectively. In this limit the spin noise spectrum is determined by
the probing sensitivity. Indeed, if spin relaxation processes and
spread of spin precession frequencies due to nuclear fields are
neglected, then for each fluctuation frequency $\omega$ there is
just one ``resonant'' Larmor frequency $\Omega_{\bm B} = \omega$ and,
due to the relation~(\ref{AC}), the corresponding trion transition
frequency $\omega_0(\omega) = (\omega - C)/A$. The intensity of the Faraday
rotation and ellipticity fluctuations at the frequency $\omega$ is
given in this limit simply by the sensitivity of the corresponding
spin signal at the optical frequency $\omega_0(\omega)$.
Interestingly, in this case the
fluctuations spectrum  of the Faraday rotation angle becomes zero at
the frequency
$A\omega_{\rm pr} +C$. The correlation between electron
  $g$-factor and optical transition energy gives rise also to the
  peculiar temporal behavior of the Faraday rotation signal.\cite{glazov2010a}

\begin{figure}[tb]
\includegraphics[width=0.45\textwidth]{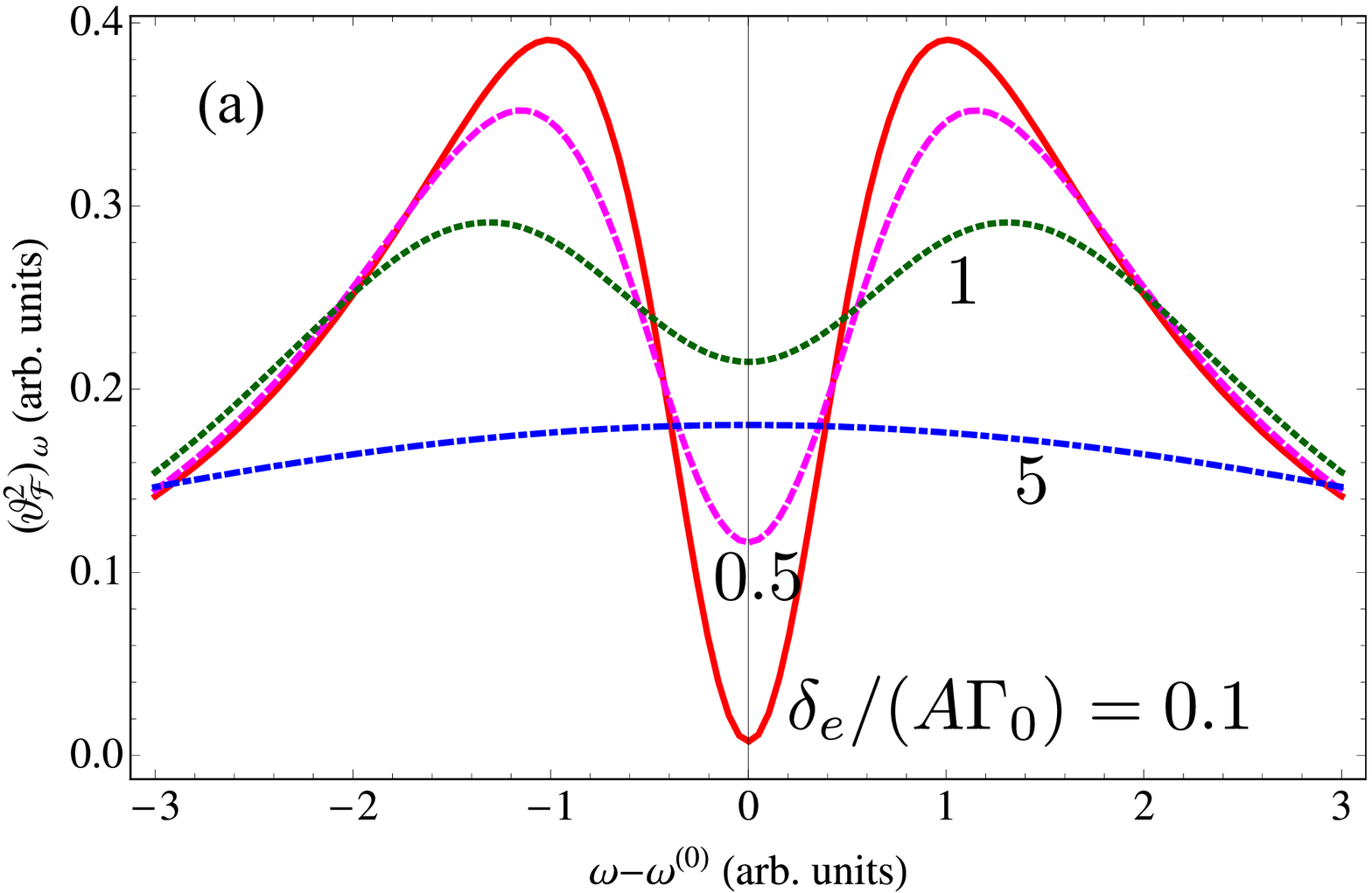}\\
\includegraphics[width=0.45\textwidth]{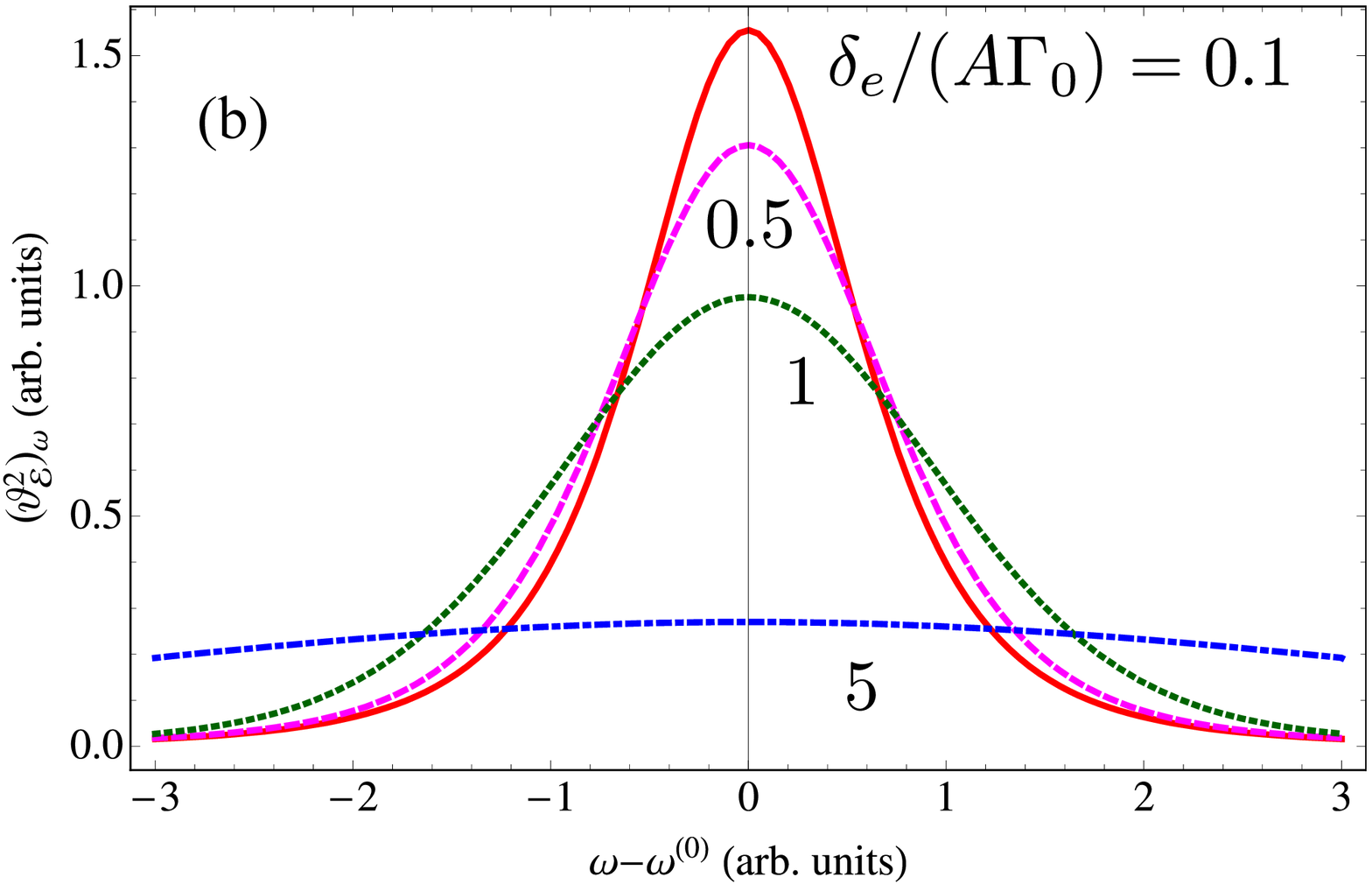}
\caption{Faraday rotation (a) and ellipticity (b) fluctuation
  spectra calculated for four
  different ratios $\delta/(A\Gamma_0)=0.1$, $0.5$, $1$ and $5$. The
  origin of frequencies is taken at the precession frequency $\omega^{(0)}=A\omega_{\rm pr}+
  C$ where the Faraday
  rotation fluctuations vanish, see text for details.}
\label{fig:sns:inhom}
\end{figure}

To model the crossover between these two limits we take a combined
distribution of electron spin precession and optical frequencies in
the form 
\[
p(\Omega,\omega_0) = \frac{1}{\sqrt{\pi}\delta_e}
\exp{\left(-\frac{(\Omega - A\omega_0 - C)^2}{\delta_e^2}\right)}p_0(\omega_0),
\]
where the function $p(\omega_0)$ describes the distribution of optical resonance frequencies and is, hereafter, taken to be flat within the frequency range of interest.
The Faraday rotation and ellipticity fluctuation spectra are shown in
Fig.~\ref{fig:sns:inhom}, panels (a) and (b), respectively. 

One can clearly see from Fig.~\ref{fig:sns:inhom} that, with an
increase of the spread of electron spin precession frequencies
controlled by the parameter  $\delta_e/(A\Gamma_0)$, the Faraday
rotation spectrum transforms from
two-maxima shape with vanishing signal in the middle to the flat
spectrum. At the same
time, the fluctuations spectrum of ellipticity signal
$(\vartheta_{\mathcal E}^2)_\omega$ simply widens with an
increase of $\delta_e/(A\Gamma_0)$. 


\section{Conclusions}

To conclude, we have developed a microscopic theory of
  electron or hole spin fluctuations in semiconductor quantum dot
  ensembles. The spin noise spectra are calculated with allowance for
  the hyperfine or dipole-dipole interaction of the charge carrier spin
  with lattice nuclei, the Zeeman effect of external magnetic field and
  inhomogeneous broadening of the electron $g$-factor. The
  spin noise features related with the spin relaxation and spin
  decoherence caused by nuclei are identified. 

The fluctuation spectra of spin-Faraday and ellipticity
  effects have been analyzed as well. It is demonstrated that their shape may
  be strongly affected by the correlation between the optical
  transition frequency and the electron spin precession frequency.

\acknowledgements

We thank A. Greilich, J.~H\"{u}bner, G.G. Kozlov, M. Oestreich, D.R. Yakovlev,
I.A. Yugova, and V.S. Zapasskii  for valuable 
discussions. Financial support of RFBR, RF President Grant
NSh-5442.2012.2, and EU projects SPANGL4Q, Spinoptronics and POLAPHEN is
gratefully acknowledged.

\appendix

\section{Spin fluctuations in the framework of the
  fluctuation-dissipation theorem}

In the framework of the linear response theory\cite{LandauStat} the electron spin
fluctuations can be related with the generalized spin
susceptibility $\mu_{\alpha\beta}$ describing the linear response of
electron spin to the generalized forces $f_\alpha$:
\begin{equation}
\label{mu:def}
\delta s_{\alpha,\omega} = \sum_\beta \mu_{\alpha\beta}(\omega)
f_{\beta,\omega}\:. 
\end{equation}
In this second description of spin fluctuations equivalent to the description provided by Eqs.~(\ref{field}), (\ref{correlator_force})
and (\ref{fd}), the force ${\bm f}(t)$ acts as a perturbation to the spin Hamiltonian~\cite{LandauStat}
\begin{equation}
\label{pert}
\hat V = - \sum_\alpha \hat{s}_\alpha f_\alpha\:,
\end{equation} 
where $\hat{s}_\alpha$ are the electron spin operators, and can be
related to the components of electron spin precession frequency in a
random magnetic field as ${\bm \Omega}_{\sim} = - {\bm f}/\hbar$. 

In the presence of the static magnetic field characterized by the spin
precession frequency $\bm \Omega$, see Eq.~(\ref{Omega:inB}), the
magnetic susceptibility ${\bm \mu}(\omega)$ can be found from the kinetic equation for the
electron spin $\bm s = \bar{\bm s} + \delta \bm s$, where
${\bar{\bm s} \equiv} \bar{\bm
s}({\bm \Omega})$ is the equilibrium spin orientation in the static 
field with the Larmor frequency ${\bm \Omega}$ and $\delta \bm s$ is the non-equilibrium spin polarization
induced by the weak fluctuating force ${\bm f}$ ($|\bm f| \ll \hbar |\bm \Omega|$): 
\begin{equation}
  \label{kinetic:gen:0}
\frac{\partial \bm s}{\partial t} + \left[\bm s \times \left( {\bm \Omega} + {\bm \Omega}_{\sim} \right) \right]  + Q\{ \bm s \}=0,
\end{equation}
where the collision integral of the form\cite{slicht}
\begin{equation}
\label{collision}
 Q\{ \bm s \} = \frac{\bm s - \bm \bar{\bm s} \left(\bm \Omega + {\bm \Omega}_{\sim} \right)}{\tau_s} 
\end{equation}
takes into account the spin relaxation to its equilibrium value for the
total field with the Larmor frequency ${\bm \Omega} + {\bm \Omega}_{\sim}$. Assuming that the Zeeman
splitting induced by the external field $\hbar\bm\Omega$ is much smaller
than the temperature of the system expressed in the units of energy,
$k_B T$, one has
\begin{equation}
\bar{\bm s}({\bm \Omega}) = - \frac{\hbar{\bm \Omega}}{4k_B T}\:,\:
\bar{\bm s}({\bm \Omega} + {\bm \Omega}_{\sim}) = - \frac{\hbar{\bm \Omega} - \bm f}{4k_B T}
\:,
\end{equation}
where $k_B$ is the Boltzmann constant.
Hence, the fluctuation $\delta \bm s$ satisfies the
following linearized equation:
\begin{equation}
\label{kinetic:gen}
-\mathrm i \omega \delta \bm s + \delta \bm s \times \bm \Omega +
\frac{\delta \bm s}{\tau_s} = \frac{1}{4k_BT} \left(\frac{\bm f}{\tau_s} -
 \bm \Omega \times \bm f \right)\:.
\end{equation}
It can be solved by using
Eqs.~(\ref{solution}) and (\ref{componchi}) 
yielding the spin susceptibility in the form
\begin{equation}
  \label{mu:res}
  \mu_{\alpha\beta} = \frac{1}{4k_B T\tau_s}
  \left[\chi_{\alpha\beta}(\omega) - \sum_{\mu\nu} \chi_{\alpha\mu}(\omega)\delta_{\mu\nu\beta}\Omega_\nu \right]\:.
\end{equation}
One can readily check that the spin noise spectral functions of Eq.~(\ref{fd}), in agreement with the general
theory,\cite{LandauStat} are expressed via the susceptibility
$\mu_{\alpha\beta}(\omega)$ as [cf. Ref.~\onlinecite{kos2010}]
\begin{equation}
\label{fdt}
(\delta s_\alpha \delta s_\beta)_\omega = \frac{\mathrm i k_B T}{\omega}
\left[\mu_{\beta\alpha}^*(\omega) - \mu_{\alpha\beta}(\omega) \right]\:.
\end{equation}
This equivalence can be proved by applying the following identity relating bilinear and linear components of the tensor ${\bm \chi}$:
\begin{eqnarray} \label{eq:useful}
\sum_{\gamma} \chi_{\alpha\gamma}(\omega) \chi_{\beta\gamma}^*(\omega) = \frac{\mathrm i}{2\omega} ~
\left\{ \chi_{\beta\alpha}^*(\omega) - \chi_{\alpha\beta}(\omega) \right.\\
 + \sum_{\mu\nu}~\left. [ \chi_{\alpha\mu}(\omega) \delta_{\mu\nu\beta} -
\chi_{\beta\mu}^*(\omega) \delta_{\mu\nu \alpha}] \Omega_\nu
\right\}\:. \nonumber 
\end{eqnarray}


\begin{thebibliography}{40}
\makeatletter
\providecommand \@ifxundefined [1]{%
 \ifx #1\undefined \expandafter \@firstoftwo
 \else \expandafter \@secondoftwo
\fi
}%
\providecommand \@ifnum [1]{%
 \ifnum #1\expandafter \@firstoftwo
 \else \expandafter \@secondoftwo
\fi
}%
\providecommand \enquote [1]{``#1''}%
\providecommand \bibnamefont  [1]{#1}%
\providecommand \bibfnamefont [1]{#1}%
\providecommand \citenamefont [1]{#1}%
\providecommand\href[0]{\@sanitize\@href}%
\providecommand\@href[1]{\endgroup\@@startlink{#1}\endgroup\@@href}%
\providecommand\@@href[1]{#1\@@endlink}%
\providecommand \@sanitize [0]{\begingroup\catcode`\&12\catcode`\#12\relax}%
\@ifxundefined \pdfoutput {\@firstoftwo}{%
 \@ifnum{\z@=\pdfoutput}{\@firstoftwo}{\@secondoftwo}%
}{%
 \providecommand\@@startlink[1]{\leavevmode}%
 \providecommand\@@endlink[0]{}%
}{%
 \providecommand\@@startlink[1]{%
  \leavevmode
  \pdfstartlink
   attr{/Border[0 0 1 ]/H/I/C[0 1 1]}%
   user{/Subtype/Link/A<</Type/Action/S/URI/URI(#1)>>}%
  \relax
 }%
 \providecommand\@@endlink[0]{\pdfendlink}%
}%
\providecommand \url  [0]{\begingroup\@sanitize \@url }%
\providecommand \@url [1]{\endgroup\@href {#1}{\urlprefix}}%
\providecommand \urlprefix [0]{URL }%
\providecommand \Eprint[0]{\href }%
\@ifxundefined \urlstyle {%
  \providecommand \doi [1]{doi:\discretionary{}{}{}#1}%
}{%
  \providecommand \doi [0]{doi:\discretionary{}{}{}\begingroup
  \urlstyle{rm}\Url }%
}%
\providecommand \doibase [0]{http://dx.doi.org/}%
\providecommand \Doi[1]{\href{\doibase#1}}%
\providecommand \bibAnnote [3]{%
  \BibitemShut{#1}%
  \begin{quotation}\noindent
    \textsc{Key:}\ #2\\\textsc{Annotation:}\ #3%
  \end{quotation}%
}%
\providecommand \bibAnnoteFile [2]{%
  \IfFileExists{#2}{\bibAnnote {#1} {#2} {\input{#2}}}{}%
}%
\providecommand \typeout [0]{\immediate \write \m@ne }%
\providecommand \selectlanguage [0]{\@gobble}%
\providecommand \bibinfo [0]{\@secondoftwo}%
\providecommand \bibfield [0]{\@secondoftwo}%
\providecommand \translation [1]{[#1]}%
\providecommand \BibitemOpen[0]{}%
\providecommand \bibitemStop [0]{}%
\providecommand \bibitemNoStop [0]{.\EOS\space}%
\providecommand \EOS [0]{\spacefactor3000\relax}%
\providecommand \BibitemShut [1]{\csname bibitem#1\endcsname}%

\bibitem{Mueller2010}%
  \BibitemOpen
  \bibfield{author}{%
  \bibinfo {author} {\bibfnamefont{G.~M.}\ \bibnamefont{M\"{u}ller}}, \bibinfo
  {author} {\bibfnamefont{M.}~\bibnamefont{Oestreich}}, \bibinfo {author}
  {\bibfnamefont{M.}~\bibnamefont{R\"{o}mer}},\ and\ \bibinfo {author}
  {\bibfnamefont{J.}~\bibnamefont{H\"{u}bner}},\ }%
  \bibfield{journal}{%
{\bibinfo {journal} {Physica E}}\ }%
  \textbf{\bibinfo {volume} {43}},\ \bibinfo {pages} {569} (\bibinfo {year}
  {2010}).

\bibitem{Crooker_Noise}%
  \BibitemOpen
  \bibfield{author}{%
  \bibinfo {author} {\bibfnamefont{S.~A.}\ \bibnamefont{Crooker}}, \bibinfo
  {author} {\bibfnamefont{D.~G.}\ \bibnamefont{Rickel}}, \bibinfo {author}
  {\bibfnamefont{A.~V.}\ \bibnamefont{Balatsky}},\ and\ \bibinfo {author}
  {\bibfnamefont{D.~L.}\ \bibnamefont{Smith}},\ }%
  \bibfield{journal}{%
  \bibinfo {journal} {Nature}\ }%
  \textbf{\bibinfo {volume} {431}},\ \bibinfo {pages} {49} (\bibinfo {year}
  {2004}).%
  \bibAnnoteFile{NoStop}{Crooker_Noise}%
\bibitem{Oestreich_noise}%
  \BibitemOpen
  \bibfield{author}{%
  \bibinfo {author} {\bibfnamefont{M.}~\bibnamefont{Oestreich}}, \bibinfo
  {author} {\bibfnamefont{M.}~\bibnamefont{R\"omer}}, \bibinfo {author}
  {\bibfnamefont{R.~J.}\ \bibnamefont{Haug}},\ and\ \bibinfo {author}
  {\bibfnamefont{D.}~\bibnamefont{H\"agele}},\ }%
  \bibfield{journal}{%
  \bibinfo {journal} {Phys. Rev. Lett.}\ }%
  \textbf{\bibinfo {volume} {95}},\ \bibinfo {pages} {216603} (\bibinfo {year} {2005}).%
  \bibAnnoteFile{NoStop}{Oestreich_noise}%
\bibitem{muller:206601}%
  \BibitemOpen
  \bibfield{author}{%
  \bibinfo {author} {\bibfnamefont{G.~M.}\ \bibnamefont{M\"{u}ller}}, \bibinfo
  {author} {\bibfnamefont{M.}~\bibnamefont{R\"{o}mer}}, \bibinfo {author}
  {\bibfnamefont{D.}~\bibnamefont{Schuh}}, \bibinfo {author}
  {\bibfnamefont{W.}~\bibnamefont{Wegscheider}}, \bibinfo {author}
  {\bibfnamefont{J.}~\bibnamefont{H\"{u}bner}},\ and\ \bibinfo {author}
  {\bibfnamefont{M.}~\bibnamefont{Oestreich}},\ }%
  \bibfield{journal}{%
 {\bibinfo {journal} {Phys. Rev. Lett.}}\
  }%
  \textbf{\bibinfo {volume} {101}},\ \bibinfo {eid} {206601} (\bibinfo {year}
  {2008}).
\bibitem{PhysRevB.79.035208}%
  \BibitemOpen
  \bibfield{author}{%
  \bibinfo {author} {\bibfnamefont{S.~A.}\ \bibnamefont{Crooker}}, \bibinfo
  {author} {\bibfnamefont{L.}~\bibnamefont{Cheng}},\ and\ \bibinfo {author}
  {\bibfnamefont{D.~L.}\ \bibnamefont{Smith}},\ }%
  \bibfield{journal}{%
 {\bibinfo {journal} {Phys. Rev. B}}\ }%
  \textbf{\bibinfo {volume} {79}},\ \bibinfo {pages} {035208} (\bibinfo {year} {2009}).
\bibitem{crooker2010}%
  \BibitemOpen
  \bibfield{author}{%
  \bibinfo {author} {\bibfnamefont{S.~A.}\ \bibnamefont{Crooker}}, \bibinfo
  {author} {\bibfnamefont{J.}~\bibnamefont{Brandt}}, \bibinfo {author}
  {\bibfnamefont{C.}~\bibnamefont{Sandfort}}, \bibinfo {author}
  {\bibfnamefont{A.}~\bibnamefont{Greilich}}, \bibinfo {author}
  {\bibfnamefont{D.~R.}\ \bibnamefont{Yakovlev}}, \bibinfo {author}
  {\bibfnamefont{D.}~\bibnamefont{Reuter}}, \bibinfo {author}
  {\bibfnamefont{A.~D.}\ \bibnamefont{Wieck}},\ and\ \bibinfo {author}
  {\bibfnamefont{M.}~\bibnamefont{Bayer}},\ }%
  \bibfield{journal}{%
 {\bibinfo {journal} {Phys. Rev. Lett.}}\
  }%
  \textbf{\bibinfo {volume} {104}},\ \bibinfo {pages} {036601} (\bibinfo {year} {2010}).%
  \bibAnnoteFile{NoStop}{crooker2010}%
\bibitem{dahbashi2012} R. Dahbashi, J. H\"{u}bner, F. Berski,
  J. Wiegand, X. Marie, K. Pierz, H. W. Schumacher, and M. Oestreich,
  Appl. Phys. Lett. {\bf 100}, 031906 (2012).

\bibitem{crooker2012}
Yan Li, N. Sinitsyn, D. L. Smith, D. Reuter, A. D. Wieck,
D. R. Yakovlev, M. Bayer, S. A. Crooker, Phys. Rev. Lett. {\bf 108},
186603 (2012). 

\bibitem{aleksandrov81}%
  \BibitemOpen
  \bibfield{author}{%
  \bibinfo {author} {\bibfnamefont{E.}~\bibnamefont{Aleksandrov}}\ and\
  \bibinfo {author} {\bibfnamefont{V.}~\bibnamefont{Zapasskii}},\ }%
  \bibfield{journal}{%
  \bibinfo {journal} {Sov. Phys. JETP}\ }%
  \textbf{\bibinfo {volume} {54}},\ \bibinfo {pages} {64} (\bibinfo {year}
  {1981})  [Zh. Exp. Teor. Fiz. {\bf 81}, 132 (1981)].%
  \bibAnnoteFile{NoStop}{aleksandrov81}%
\bibitem{sherman} M. M. Glazov and E. Ya. Sherman,
  Phys. Rev. Lett. {\bf 107}, 156602 (2011).


\bibitem{A.Greilich07212006}%
  \BibitemOpen
  \bibfield{author}{%
  \bibinfo {author} {\bibfnamefont{A.}~\bibnamefont{Greilich}}, \bibinfo
  {author} {\bibfnamefont{D.~R.}\ \bibnamefont{Yakovlev}}, \bibinfo {author}
  {\bibfnamefont{A.}~\bibnamefont{Shabaev}}, \bibinfo {author}
  {\bibfnamefont{A.~L.}\ \bibnamefont{Efros}}, \bibinfo {author}
  {\bibfnamefont{I.~A.}\ \bibnamefont{Yugova}}, \bibinfo {author}
  {\bibfnamefont{R.}~\bibnamefont{Oulton}}, \bibinfo {author}
  {\bibfnamefont{V.}~\bibnamefont{Stavarache}}, \bibinfo {author}
  {\bibfnamefont{D.}~\bibnamefont{Reuter}}, \bibinfo {author}
  {\bibfnamefont{A.}~\bibnamefont{Wieck}},\ and\ \bibinfo {author}
  {\bibfnamefont{M.}~\bibnamefont{Bayer}},\ }%
  \bibfield{journal}{%
 {\bibinfo {journal} {Science}}\ }%
  \textbf{\bibinfo {volume} {313}},\ \bibinfo {pages} {341} (\bibinfo {year}
  {2006}).
\bibitem{A.Greilich09282007}%
  \BibitemOpen
  \bibfield{author}{%
  \bibinfo {author} {\bibfnamefont{A.}~\bibnamefont{Greilich}}, \bibinfo
  {author} {\bibfnamefont{A.}~\bibnamefont{Shabaev}}, \bibinfo {author}
  {\bibfnamefont{D.~R.}\ \bibnamefont{Yakovlev}}, \bibinfo {author}
  {\bibfnamefont{A.~L.}\ \bibnamefont{Efros}}, \bibinfo {author}
  {\bibfnamefont{I.~A.}\ \bibnamefont{Yugova}}, \bibinfo {author}
  {\bibfnamefont{D.}~\bibnamefont{Reuter}}, \bibinfo {author}
  {\bibfnamefont{A.~D.}\ \bibnamefont{Wieck}},\ and\ \bibinfo {author}
  {\bibfnamefont{M.}~\bibnamefont{Bayer}},\ }%
  \bibfield{journal}{%
 {\bibinfo {journal} {Science}}\ }%
  \textbf{\bibinfo {volume} {317}},\ \bibinfo {pages} {1896} (\bibinfo {year}
  {2007}).
\bibitem{yakovlev_bayer}%
  \BibitemOpen
  \bibfield{author}{%
  \bibinfo {author} {\bibfnamefont{D.}~\bibnamefont{Yakovlev}}\ and\ \bibinfo
  {author} {\bibfnamefont{M.}~\bibnamefont{Bayer}},\ }%
 in
  {\bibinfo {title} {Spin physics in semiconductors}}, Ed. 
  M. Dyakonov, Chap. 6 \ (\bibinfo
  {publisher} {Springer},\ \bibinfo {year} {2008}).
\bibitem{glazov:review}%
  \BibitemOpen
  \bibfield{author}{%
  \bibinfo {author} {\bibfnamefont{M.~M.}\ \bibnamefont{Glazov}},\ }%
  \bibfield{journal}{%
  \bibinfo {journal} {Phys. Solid State}\ }%
  \textbf{\bibinfo {volume} {54}},\ \bibinfo {pages} {1} (\bibinfo {year}
  {2012}) [Fiz. Tverd. Tela {\bf 54}, 3 (2012)].%
  \bibAnnoteFile{NoStop}{glazov:review}%
\bibitem{PhysRevB.64.125316}%
  \BibitemOpen
  \bibfield{author}{%
  \bibinfo {author} {\bibfnamefont{A.~V.}\ \bibnamefont{Khaetskii}}\ and\
  \bibinfo {author} {\bibfnamefont{Y.~V.}\ \bibnamefont{Nazarov}},\ }%
  \bibfield{journal}{%
  {\bibinfo {journal} {Phys. Rev. B}}\ }%
  \textbf{\bibinfo {volume} {64}},\ \bibinfo {pages} {125316} (\bibinfo {year} {2001}).%
  \bibAnnoteFile{NoStop}{PhysRevB.64.125316}%
\bibitem{PhysRevB.66.161318}%
  \BibitemOpen
  \bibfield{author}{%
  \bibinfo {author} {\bibfnamefont{L.~M.}\ \bibnamefont{Woods}}, \bibinfo
  {author} {\bibfnamefont{T.~L.}\ \bibnamefont{Reinecke}},\ and\ \bibinfo
  {author} {\bibfnamefont{Y.}~\bibnamefont{Lyanda-Geller}},\ }%
  \bibfield{journal}{%
  {\bibinfo {journal} {Phys. Rev. B}}\ }%
  \textbf{\bibinfo {volume} {66}},\ \bibinfo {pages} {161318} (\bibinfo {year} {2002}).%
  \bibAnnoteFile{NoStop}{PhysRevB.66.161318}%
\bibitem{kkm_nucl_book}%
  \BibitemOpen
  \bibfield{author}{%
  \bibinfo {author} {\bibfnamefont{V.}~\bibnamefont{Kalevich}}, \bibinfo
  {author} {\bibfnamefont{K.}~\bibnamefont{Kavokin}},\ and\ \bibinfo {author}
  {\bibfnamefont{I.}~\bibnamefont{Merkulov}},\ }%
in
  {\bibinfo {title} {Spin physics in semiconductors}}, Ed. by
  M. Dyakonov, \bibinfo {chapter}
  {Chap. 11} \ (\bibinfo
  {publisher} {Springer},\ \bibinfo {year} {2008}).
\bibitem{merkulov02}%
  \BibitemOpen
  \bibfield{author}{%
  \bibinfo {author} {\bibfnamefont{I.~A.}\ \bibnamefont{Merkulov}}, \bibinfo
  {author} {\bibfnamefont{A.~L.}\ \bibnamefont{Efros}},\ and\ \bibinfo {author}
  {\bibfnamefont{M.}~\bibnamefont{Rosen}},\ }%
  \bibfield{journal}{%
  \bibinfo {journal} {Phys. Rev. B}\ }%
  \textbf{\bibinfo {volume} {65}},\ \bibinfo {pages} {205309} (\bibinfo {year}
  {2002}).%
  \bibAnnoteFile{NoStop}{merkulov02}%
\bibitem{PhysRevLett.88.186802}%
  \BibitemOpen
  \bibfield{author}{%
  \bibinfo {author} {\bibfnamefont{A.~V.}\ \bibnamefont{Khaetskii}}, \bibinfo
  {author} {\bibfnamefont{D.}~\bibnamefont{Loss}},\ and\ \bibinfo {author}
  {\bibfnamefont{L.}~\bibnamefont{Glazman}},\ }%
  \bibfield{journal}{%
  {\bibinfo {journal} {Phys. Rev. Lett.}}\
  }%
  \textbf{\bibinfo {volume} {88}},\ \bibinfo {pages} {186802} (\bibinfo {year} {2002}).%
  \bibAnnoteFile{NoStop}{PhysRevLett.88.186802}%
\bibitem{yugova11}%
  \BibitemOpen
  \bibfield{author}{%
  \bibinfo {author} {\bibfnamefont{M.~M.}\ \bibnamefont{{Glazov}}}, \bibinfo
  {author} {\bibfnamefont{I.~A.}\ \bibnamefont{{Yugova}}},\ and\ \bibinfo
  {author} {\bibfnamefont{A.~L.}\ \bibnamefont{{Efros}}},\ }%
  \bibfield{journal}{%
  \bibinfo {journal} {Phys. Rev. B}\ }%
  \textbf{\bibinfo {volume} {85}},\ \bibinfo {pages} {041303} (\bibinfo {year} {2012}).%
  \bibAnnoteFile{NoStop}{yugova11}%
\bibitem{ggk} S. Gantsevich, V. Gurevich, and R. Katilius,
    Riv. Nuovo Cimento {\bf 2}, 1 (1979).
\bibitem{LandauStat} L. D. Landau and E. M. Lifshitz, Statistical Physics, Part 1 (Course of Theoretical Physics, Vol. 5), Chap. XII  (Butterworth-Heinemann, Oxford, 2000).
\bibitem{lax2} M. Lax, Rev. Mod. Phys. {\bf 38}, 541 (1966).
\bibitem{LandauKin} L. P. Pitaevskii and
    E. M. Lifshitz, Physical Kinetics, Butterworth-Heinemann (Course
    of Theoretical Physics, Vol. 10), Chap. I (Butterworth-Heinemann, Oxford, 1999).
 \bibitem{merkulov} I. A. Merkulov, G. Alvarez, D. R. Yakovlev, and
   T. C. Schulthess, 
Phys. Rev. B {\bf 81}, 115107 (2010). 


\bibitem{ivchenko73fluct_eng}%
  \BibitemOpen
  \bibfield{author}{%
  \bibinfo {author} {\bibfnamefont{E.~L.}\ \bibnamefont{Ivchenko}},\ }%
  \bibfield{journal}{%
  \bibinfo {journal} {Sov. Phys. Solid State}\ }%
  \textbf{\bibinfo {volume} {7}},\ \bibinfo {pages} {998} (\bibinfo {year}
  {1974}) [Fiz. Tverd. Tela {\bf 7}, 1489 (1974)].%
  \bibAnnoteFile{NoStop}{ivchenko73fluct_eng}%
\bibitem{coupled1} Provided that the electron spin relaxation time $\tau_s$ is much longer than the typical timescale of the nuclear field variation the coupling between electron and nuclear spin may result in slow, $1/\ln{t}$, electron spin decay (at $t<\tau_s$) and, correspondingly, $1/\omega$ feature in the spin noise spectrum (at $\omega>\tau_s^{-1}$, see Refs.~\onlinecite{coupled} for details. Experimentally, divergent low-frequency spin fluctuations were observed by Yan Li et al in Ref.~\onlinecite{crooker2012} in the presence of magnetic field.
\bibitem{coupled}R. de Sousa and S. Das
    Sarma, Phys. Rev. B {\bf 67}, 033301 (2003); K. A. Al-Hassanieh,
    V. V. Dobrovitski, E. Dagotto, and B. N. Harmon,
    Phys. Rev. Lett. {\bf 97}, 037204 (2006); R.-B. Liu, W. Yao,
    L. Sham, New Journal of Physics  {\bf 9}, 226 (2007); E. Barnes, L.
    Cywinski, and S. Das Sarma, Phys. Rev. B {\bf 84}, 155315 (2011).

\bibitem{zhang06} Wenxian Zhang, V. V. Dobrovitski,
  K. A. Al-Hassanieh, E. Dagotto, and B. N. Harmon, Phys. Rev. B {\bf
    74,} 205313 (2006).


  \bibitem{chemphys} K. Schulten and P. G. Wolynes, J. Chem. Phys. {\bf 68}, 3292 (1978). 
\bibitem{perel76eng}%
  \BibitemOpen
  \bibfield{author}{%
  \bibinfo {author} {\bibfnamefont{E.}~\bibnamefont{Gryncharova}}\ and\
  \bibinfo {author} {\bibfnamefont{V.}~\bibnamefont{Perel'}},\ }%
  \bibfield{journal}{%
  \bibinfo {journal} {Sov. Phys. Semicond.}\ }%
  \textbf{\bibinfo {volume} {11}}, 997 (\bibinfo {year} {1977})
  [Fiz. Tekhn. Polupr. {\bf 11}, 1697 (1977)].%
  \bibAnnoteFile{NoStop}{perel76eng}%
\bibitem{PhysRevB.78.155329}%
  \BibitemOpen
  \bibfield{author}{%
  \bibinfo {author} {\bibfnamefont{J.}~\bibnamefont{Fischer}}, \bibinfo
  {author} {\bibfnamefont{W.~A.}\ \bibnamefont{Coish}}, \bibinfo {author}
  {\bibfnamefont{D.~V.}\ \bibnamefont{Bulaev}},\ and\ \bibinfo {author}
  {\bibfnamefont{D.}~\bibnamefont{Loss}},\ }%
  \bibfield{journal}{%
 {\bibinfo {journal} {Phys. Rev. B}}\ }%
  \textbf{\bibinfo {volume} {78}},\ \bibinfo {pages} {155329} (\bibinfo {year} {2008}).%
  \bibAnnoteFile{NoStop}{PhysRevB.78.155329}%
\bibitem{PhysRevB.79.195440}%
  \BibitemOpen
  \bibfield{author}{%
  \bibinfo {author} {\bibfnamefont{C.}~\bibnamefont{Testelin}}, \bibinfo
  {author} {\bibfnamefont{F.}~\bibnamefont{Bernardot}}, \bibinfo {author}
  {\bibfnamefont{B.}~\bibnamefont{Eble}},\ and\ \bibinfo {author}
  {\bibfnamefont{M.}~\bibnamefont{Chamarro}},\ }%
  \bibfield{journal}{%
  {\bibinfo {journal} {Phys. Rev. B}}\ }%
  \textbf{\bibinfo {volume} {79}},\ \bibinfo {pages} {195440} (\bibinfo {year} {2009}).%
  \bibAnnoteFile{NoStop}{PhysRevB.79.195440}%
\bibitem{PhysRevLett.102.146601}%
  \BibitemOpen
  \bibfield{author}{%
  \bibinfo {author} {\bibfnamefont{B.}~\bibnamefont{Eble}}, \bibinfo {author}
  {\bibfnamefont{C.}~\bibnamefont{Testelin}}, \bibinfo {author}
  {\bibfnamefont{P.}~\bibnamefont{Desfonds}}, \bibinfo {author}
  {\bibfnamefont{F.}~\bibnamefont{Bernardot}}, \bibinfo {author}
  {\bibfnamefont{A.}~\bibnamefont{Balocchi}}, \bibinfo {author}
  {\bibfnamefont{T.}~\bibnamefont{Amand}}, \bibinfo {author}
  {\bibfnamefont{A.}~\bibnamefont{Miard}}, \bibinfo {author}
  {\bibfnamefont{A.}~\bibnamefont{Lema\^\i{}tre}}, \bibinfo {author}
  {\bibfnamefont{X.}~\bibnamefont{Marie}},\ and\ \bibinfo {author}
  {\bibfnamefont{M.}~\bibnamefont{Chamarro}},\ }%
  \bibfield{journal}{%
  {\bibinfo {journal} {Phys. Rev. Lett.}}\
  }%
  \textbf{\bibinfo {volume} {102}},\ \bibinfo {pages} {146601} (\bibinfo {year} {2009}).%
  \bibAnnoteFile{NoStop}{PhysRevLett.102.146601}%
\bibitem{braun} M. Braun and J. K\"{o}nig, Phys.Rev. B {\bf 75} 085310 (2007).

\bibitem{glazov08a}%
  \BibitemOpen
  \bibfield{author}{%
  \bibinfo {author} {\bibfnamefont{M.~M.}\ \bibnamefont{Glazov}}\ and\ \bibinfo
  {author} {\bibfnamefont{E.~L.}\ \bibnamefont{Ivchenko}},\ }%
  \bibfield{journal}{%
  \bibinfo {journal} {Semiconductors}\ }%
  \textbf{\bibinfo {volume} {42}},\ \bibinfo {pages} {951} (\bibinfo {year}
  {2008}) [Fiz. Tekhn. Polupr. {\bf 42}, 966 (2008)].%
  \bibAnnoteFile{NoStop}{glazov08a}%
\bibitem{yugova09}%
  \BibitemOpen
  \bibfield{author}{%
  \bibinfo {author} {\bibfnamefont{I.~A.}\ \bibnamefont{Yugova}}, \bibinfo
  {author} {\bibfnamefont{M.~M.}\ \bibnamefont{Glazov}}, \bibinfo {author}
  {\bibfnamefont{E.~L.}\ \bibnamefont{Ivchenko}},\ and\ \bibinfo {author}
  {\bibfnamefont{A.~L.}\ \bibnamefont{Efros}},\ }%
  \bibfield{journal}{%
 {\bibinfo {journal} {Phys. Rev. B}}\ }%
  \textbf{\bibinfo {volume} {80}},\ \bibinfo {eid} {104436} (\bibinfo {year}
  {2009}).

\bibitem{ivchenko05a} E. L. Ivchenko, Optical Spectroscopy of
  Semiconductor Nanostructures (Alpha Science, Harrow, UK 2005).

\bibitem{glazov2010a}%
  \BibitemOpen
  \bibfield{author}{%
  \bibinfo {author} {\bibfnamefont{M.~M.}\ \bibnamefont{Glazov}}, \bibinfo
  {author} {\bibfnamefont{I.~A.}\ \bibnamefont{Yugova}}, \bibinfo {author}
  {\bibfnamefont{S.}~\bibnamefont{Spatzek}}, \bibinfo {author}
  {\bibfnamefont{A.}~\bibnamefont{Schwan}}, \bibinfo {author}
  {\bibfnamefont{S.}~\bibnamefont{Varwig}}, \bibinfo {author}
  {\bibfnamefont{D.~R.}\ \bibnamefont{Yakovlev}}, \bibinfo {author}
  {\bibfnamefont{D.}~\bibnamefont{Reuter}}, \bibinfo {author}
  {\bibfnamefont{A.~D.}\ \bibnamefont{Wieck}},\ and\ \bibinfo {author}
  {\bibfnamefont{M.}~\bibnamefont{Bayer}},\ }%
  \bibfield{journal}{%
 {\bibinfo {journal} {Phys. Rev. B}}\ }%
  \textbf{\bibinfo {volume} {82}},\ \bibinfo {pages} {155325} (\bibinfo {year} {2010}).%
  \bibAnnoteFile{NoStop}{glazov2010a}%
\bibitem{kos2010}%
  \BibitemOpen
  \bibfield{author}{%
  \bibinfo {author} {\bibfnamefont{\v{S}.}\ \bibnamefont{Kos}}, \bibinfo
  {author} {\bibfnamefont{A.~V.}\ \bibnamefont{Balatsky}}, \bibinfo {author}
  {\bibfnamefont{P.~B.}\ \bibnamefont{Littlewood}},\ and\ \bibinfo {author}
  {\bibfnamefont{D.~L.}\ \bibnamefont{Smith}},\ }%
  \bibfield{journal}{%
 {\bibinfo {journal} {Phys. Rev. B}}\ }%
  \textbf{\bibinfo {volume} {81}},\ \bibinfo {pages} {064407} (\bibinfo {year} {2010}).%
  \bibAnnoteFile{NoStop}{kos2010}%
  \bibitem{slicht} See, e.g., C.P. Slichter, Principles of Magnetic Resonance, Chap. II, V (Springer-Verlag, Berlin, 1980).
\end{thebibliography}
\end{document}